\documentclass[aps,pra,reprint,showpacs,superscriptaddress,longbibliography]{revtex4-2} %
\usepackage{mathtools,amssymb,graphicx,units}
\usepackage[usenames,dvipsnames]{color}
\usepackage[plainpages=false,pdfpagelabels,colorlinks=true,linkcolor=red,urlcolor=blue,citecolor=blue,pdftitle={Title},pdfauthor={},pdfdisplaydoctitle=true,pdfduplex=DuplexFlipLongEdge]{hyperref}
\usepackage{verbatim}
\usepackage[english]{babel}

\usepackage[normalem]{ulem}
\usepackage{appendix}
\usepackage{bbm}
\usepackage{makecell}

\newcommand{\beginsupplement}{%
        \setcounter{table}{0}
        \renewcommand{\thetable}{S\arabic{table}}%
        \setcounter{figure}{0}
        \renewcommand{\thefigure}{S\arabic{figure}}%
}
\graphicspath{{./fig/}}

\begin{document}
\title{Reentrant localization transition induced by a composite potential}
\author{Xingbo Wei}
\email{weixingbo@zstu.edu.cn}
\affiliation{Zhejiang Key Laboratory of Quantum State Control and Optical Field Manipulation, Department of Physics, Zhejiang Sci-Tech University, 310018 Hangzhou, China}

\nocite{data}
\begin{abstract}
We numerically investigate the localization transition in a one-dimensional system subjected to a composite potential consisting of periodic and quasi-periodic components. For the rational wave vector $\alpha=1/2$, the periodic component reduces to a staggered potential, which has been reported to induce the reentrant localization transition. In addition to $\alpha=1/2$, we find that other rational wave vectors can also lead to the reentrant phenomenon. To investigate the underlying mechanisms of the reentrant localization transition, we vary the parameters of the composite potential and find that the reentrant localization transition is sensitive to the phase factor of the periodic component. By further studying the structure of the periodic component, we confirm that this sensitivity arises from the periodic phase factor modulating the mirror symmetry. Finally, we map out a global phase diagram and reveal that the reentrant localization transition originates from a paradoxical effect: increasing the amplitude of the periodic component enhances localization but simultaneously strengthens the mirror symmetry, which favors the formation of extended states. Our numerical analysis
suggests that the interplay between these competing factors drives the reentrant localization transition.	
\end{abstract}

\maketitle
\section{Introduction}
Anderson localization refers to the absence of transport diffusion due to the disorder effect, which has been extensively studied since 1958~\cite{Anderson1492}. According to the scaling theory, all states in one- and two-dimensional systems with disordered potentials lacking symmetry are localized~\cite{Abrahams673,Lee882}. In contrast, three-dimensional systems exhibit a transition from an extended phase to the Anderson localized phase as the disorder amplitude increases~\cite{Abrahams673}. During the transition, the spectrum manifests a coexistence of extended and localized states, separated by a critical energy, dubbed the mobility edge~\cite{Mott1265}. Violating the scaling theory, studies of correlated disorder show that the extended-localized transition can also occur in one-dimensional systems~\cite{aubry1980,Dunlap88}. A well-known example is the Aubry-Andr{\'e} (AA) model, a quasi-periodic system with an exactly solvable localization transition point~\cite{aubry1980}. Quasi-periodic systems possess unique properties distinct from those of disordered and periodic systems, providing a rich playground for studying intriguing physics, e.g., mobility edges~\cite{Ganeshan146601,Li085119}, critical states~\cite{Wang073204,Wang104504}, and nontrivial topological phases~\cite{Cai176403,Madeira224505}.

Generally, as the amplitude of the quasi-periodic potential increases, the system undergoes a single localization transition~\cite{aubry1980,Ganeshan146601,Li085119}. However, in certain cases, multiple localization transitions can occur, a phenomenon known as the reentrant localization transition~\cite{PhysRevLett622714,Kraus116404,PhysRevResearch3033257,goblot2020emergence,Tabanelli184208,RoyChain,Roy214203,Padhanlattice, Qi224201}. A prominent example is the interpolating Aubry-Andr{\'e}-Fibonacci model~\cite{Kraus116404,PhysRevResearch3033257,goblot2020emergence,Tabanelli184208}. Recently, this phenomenon was observed in a one-dimensional system with constrained quasi-periodic potentials~\cite{RoyChain,Roy214203}, where two distinct localization transitions occur. Further studies demonstrate that the reentrant localization transition can be realized without any constraints, simply by combining quasi-periodic and staggered onsite potentials~\cite{Padhanlattice, Qi224201}. Since its discovery~\cite{PhysRevLett622714}, the reentrant localization transition has been extensively studied across various models, including the Su-Schrieffer-Heeger model with random-dimer disorder~\cite{Zuo013305,xu2024observation} and the modified Rosenzweig-Porter model~\cite{ghosh2024ization}. Furthermore, this phenomenon is also predicted in a range of systems, e.g., spinful~\cite{Guan033305,sarkar2024ollisional}, high-dimensional~\cite{Xiaolong116,Queiroz214202}, and non-Hermitian systems~\cite{Wang075128,Han054204,Chaohuaac430b,Zhou054307,PadhanL020203,jiang2021mobility}. Alongside theoretical studies, experimental investigations into the reentrant localization transition also progress rapidly. It has been realized in the random-dimer disordered SSH photonic lattice~\cite{xu2024observation}, photonic quasicrystals~\cite{Vaidya033170}, and polaritonic wires~\cite{goblot2020emergence}. 

In this work, we numerically study the reentrant localization transition using a composite potential consisting of periodic and quasi-periodic components. Unlike previous studies that use the staggered modulation~\cite{Padhanlattice, Qi224201, Zuo013305, xu2024observation, li2023multishift}, we adopt a more general form of the periodic modulation. This modification allows us to address a fundamental question: Can periodic modulations beyond the staggered type induce the reentrant localization transition? By analyzing key parameters of the periodic component, such as the amplitude, period, and phase factor, we aim to uncover the role of the periodic modulation in the formation of the reentrant localization transition. More importantly, this investigation provides insights into the fundamental mechanisms of the reentrant localization transition driven by the composite potential. These findings offer valuable guidance for engineering and controlling reentrant phenomena.

\section{Model}
We numerically investigate a one-dimensional lattice model with $L$ sites and periodic boundary conditions, described by the following Hamiltonian, 
\begin{equation}\label{hub}
	\begin{aligned}
	\mathcal{\hat{H}}=&-t\sum_{j=1}^{L} (\hat{c}_j^{\dagger} \hat{c}_{j+1}+ h.c.) \\
	&+V\sum_{j=1}^{L} \cos (2 \pi \alpha j+\phi_v)\hat{n}_j +\lambda \sum_{j=1}^{L} \cos (2 \pi \beta j+\phi_{\lambda}) \hat{n}_j,
	\end{aligned}
\end{equation}
where $\hat{c}_j^{\dagger} (\hat{c}_j)$ is the fermionic creation (annihilation) operator at lattice site $j$, and $\hat{n}_j=\hat{c}_j^{\dagger} \hat{c}_j$ represents the particle number operator. $t\equiv1$ denotes the nearest-neighbor hopping strength. $V$ and $\lambda$ are the amplitudes of the periodic and quasi-periodic potentials, respectively. The periodic potential is characterized by a rational wave-vector $\alpha$, whereas the quasi-periodic potential is modulated by an irrational wave-vector $\beta$. Unless otherwise specified, we set $\alpha={1}/{9}$ and $\beta=(\sqrt{5}-1)/2$. Other situations of different $\alpha$ are discussed in Appendix~\ref{add1}. $\phi_v$ ($\phi_\lambda$) is the phase factor of the periodic (quasi-periodic) potential.

For $V\neq0$ and $\lambda=0$, the potential exhibits discrete translational symmetry and the Hamiltonian commutes with the translation operator $[\mathcal{\hat{H}}, \mathcal{\hat{T}}]=0$. According to the Bloch's theorem, all eigenstates are Bloch waves, showing extension in space~\cite{Abanin021001}. For $V=0$ and $\lambda\neq0$, Eq.~\eqref{hub} reduces to the AA model~\cite{aubry1980}. The system undergoes a transition from the extended phase to the Anderson localized phase as $\lambda$ increases and the transition point is at $\lambda/t=2$~\cite{aubry1980}. For $V, \lambda\neq0$ and $\alpha=1/2$, the periodic potential is staggered and multiple localization transitions are observed by regulating $V$ and $\lambda$~\cite{Padhanlattice}. 

In this work, we use multiple quantities to characterize the localization transition. The first one is the fractal dimension $D=-\ln(\mathcal{I})/\ln(L)$~\cite{Mac180601,RevModPhys801355}, which is associated with the inverse participation ratio (IPR) $\mathcal{I}=\sum_j|\psi_j|^{4}$, where $\psi_j$ is the eigenstate at site $j$. The fractal dimension is used to characterize different states: $D=1$ for extended or ergodic states, $D=0$ for localized states, and $0<D<1$ for critical states~\cite{Roy214203,Liu012222,Qi224201,Yucheng053312}. The second quantity is the Lyapunov exponent $\gamma$, which is widely used to distinguish localized states from other states. For localized states $\gamma>0$, whereas for extended states or critical states, $\gamma=0$~\cite{Zhou176401,LiL041102,Fuming174501,wang2024emergent}. The details of the Lyapunov exponent are provided in Appendix~\ref{add2}.

\begin{figure}[htbp]
	\includegraphics[width=1.0\columnwidth,height=0.8\columnwidth]{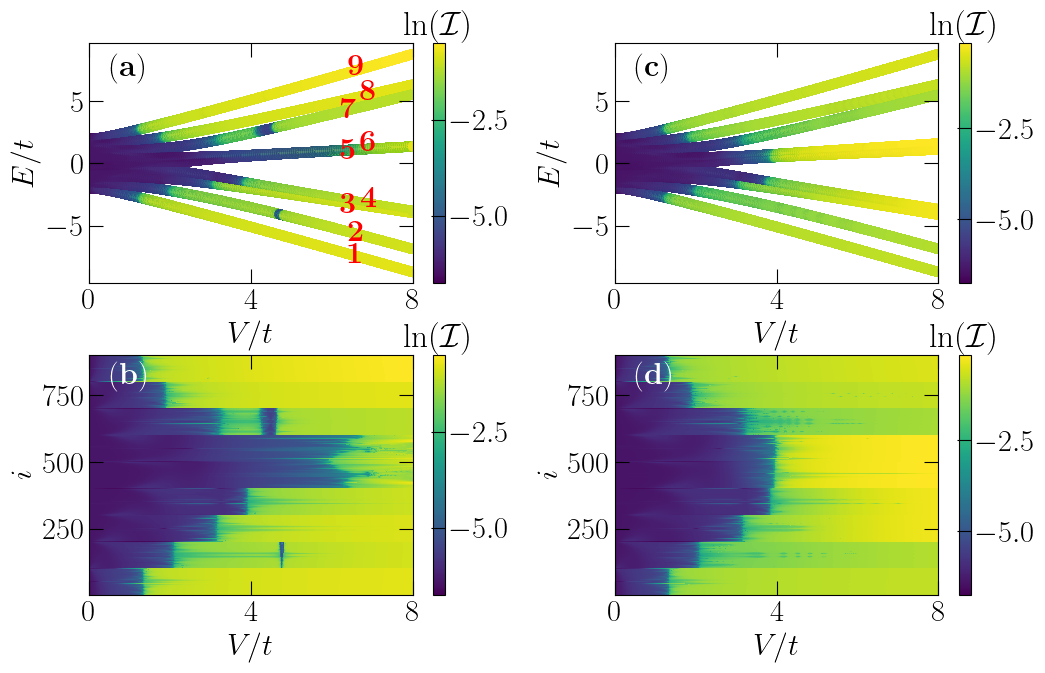}
	\vspace{-0.4cm}
	\caption{The spectra as a function of $V/t$ for $\phi_v=0$ (left panels) and $\phi_v=0.01\pi$ (right panels). The ordinates in (a) and (c) represent eigenvalues, while those in (b) and (d) denote the energy level index. $L=900$, $\lambda/t=0.007$, and $\phi_\lambda=0$. The colors represent the values of $\ln(\mathcal{I})$. In (a), the red numbers represent the energy band index.
	}
	\label{fig1}
\end{figure}

\section{Results}
In Fig.~\ref{fig1} (a), we show the spectrum as a function of $V/t$ for the periodic phase factor $\phi_v=0$. Consistent with previous studies~\cite{RoyChain,Roy214203,Padhanlattice, Qi224201,ganguly2024phenomenon,Aditya035402}, distinct mobility edges are observed in the spectrum, where extended states are distributed in the middle of the spectrum, while localized states occupy both sides. As $V/t$ increases, the spectrum splits into nine energy bands, corresponding to the rational wave vector $\alpha=1/9$. This is because the spectrum structure is determined by the periodic potential for $V\gg t,\lambda$.
From low energy to high energy, reentrant localization transitions occur in the second and seventh energy bands marked by red numbers. In these two energy bands, the system enters the localized phase twice as $V/t$ increases. This phenomenon is unique to these two bands because they form narrow bands with a high density of states, which is conducive to the overlap of localized states to form extended states, as shown in Appendix~\ref{add4}. It is worth noting that here we adopt a generalized definition of the reentrant localization transition, referring to scenarios where a specific energy band undergoes localization transitions twice or more~\cite{Vaidya033170,Tabanelli184208}. A stricter definition, based on the entire spectrum, requires that the localized phase during the transition satisfies the condition that all states in the spectrum are localized~\cite{RoyChain,Roy214203,Padhanlattice}, which is also implemented in this work.
In Fig.~\ref{fig1} (b), we use the energy level index as the ordinate, in which nine energy bands are clearly visible, and each band has $L/9$ energy levels. In Appendix~\ref{add1}, we also use the same coordinates to show the spectra for $\alpha=1/3, 1/4, 1/5, 1/6, 1/100$, and they all confirm the existence of the reentrant localization transition. It suggests that the reentrant localization transition is a widespread phenomenon in systems governed by composite potentials. In Fig.~\ref{fig1} (c) and (d), we adjust the periodic phase factor $\phi_v$ slightly, which causes subtle changes in the spectrum structure. However, compared to Fig.~\ref{fig1} (a) and (b), a slight change in $\phi_v$ significantly shifts the localization transition points in the fifth and sixth energy bands. More importantly, the reentrant localization transitions in the second and seventh energy bands disappear. It implies that the phase factor of the periodic potential plays a critical role in the formation of the reentrant localization transition.        

\begin{figure}[htbp]	\includegraphics[width=1.0\columnwidth,height=0.8\columnwidth]{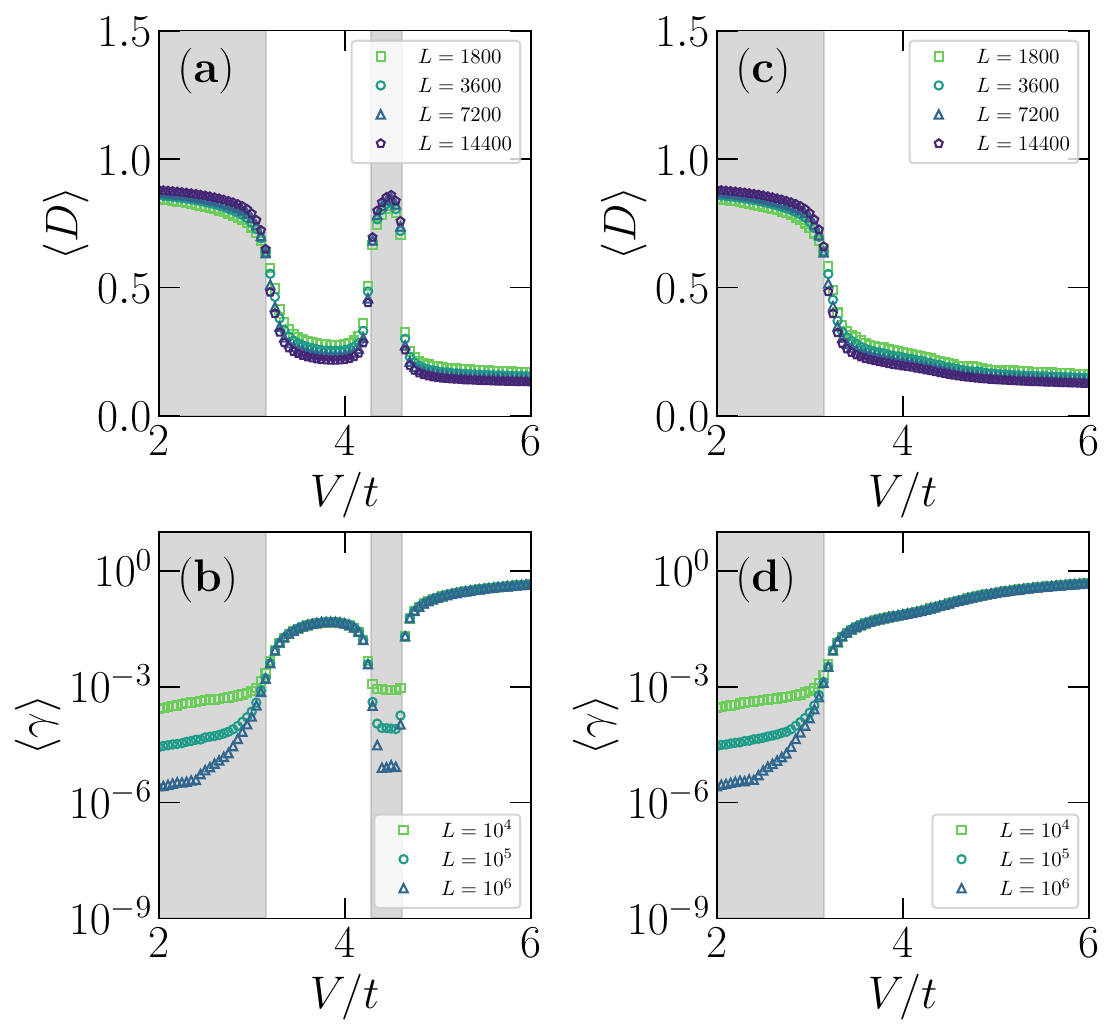}
	\vspace{-0.4cm}
	\caption{Mean fractal dimension $\langle D \rangle$ and mean Lyapunov exponent $\langle \gamma \rangle$ as a function of $V/t$ for $\phi_v=0$ (left panels) and $\phi_v=0.01\pi$ (right panels). $\lambda/t=0.007$ and $\phi_\lambda=0$. We select 100 target energies in the seventh energy band to calculate the mean Lyapunov exponent. The gray-shaded regions indicate the extended phases.
		}
	\label{fig2}
\end{figure}

To clearly demonstrate the reentrant phenomenon, we select the seventh energy band as an example for detailed analysis. In Fig.~\ref{fig2} (a), we show the mean fractal dimension of the seventh energy band as a function of $V/t$. As $V/t$ increases, the system undergoes a sequence of extended-localized-extended-localized phases. In the extended phases, highlighted by gray-shaded regions, the mean fractal dimension $\langle D \rangle$ increases with the system size. In the thermodynamic limit, $\langle D \rangle$ is expected to approach $\langle D \rangle=1$~\cite{Roy214203,Liu012222,Qi224201,Yucheng053312}. Conversely, in the localized phases, $\langle D \rangle$ decreases as the system size grows and approaches $\langle D \rangle=0$. In Appendix~\ref{add3}, we perform the finite-size scaling to further analyze the critical amplitudes and critical exponents by using the ansatz $f[(V-V_c)L^{1/\nu}]$~\cite{Luitz081103,Lv013315}. According to the scaling analysis, we determine that the system first enters the localized phase at $V/t=3.15$. With a further increase in $V/t$, the localized phase vanishes and the system transitions into the extended phase at $V/t=4.28$. Finally, the system re-enters the localized phase at $V/t=4.62$. The critical exponents are found to be $\nu \approx 3.3$, $3.5$, and $3.9$, consistent with the Harris criterion $\nu > 2/d$~\cite{harris1974effect}, where $d$ represents the system dimension. To rule out the influence of finite-size effects on the reentrant localization transition, we calculate the mean Lyapunov exponent $\langle \gamma \rangle$ for larger system sizes $L=10^4, 10^5, 10^6$ in Fig.~\ref{fig2} (b). The mean Lyapunov exponent remains constant in the localized phases, whereas it decreases with the size in the extended phases and approaches zero in the thermodynamic limit~\cite{Zhou176401,LiL041102}. The transition points indicated by the mean Lyapunov exponent are consistent with those identified using the mean fractal dimension in Fig.~\ref{fig2} (a). By slightly adjusting the phase factor of the periodic potential, the extended phase within the interval $4.28<V/t<4.62$ disappears, as shown in Fig.~\ref{fig2} (c) and (d). Remarkably, this adjustment has minimal impact on the first localization transition, with the transition point remaining at $V/t \approx 3.15$. After this localization transition, the mean fractal dimension decreases as $V/t$ increases, while the mean Lyapunov exponent increases. These trends indicate that the increasing periodic potential amplitude $V/t$ monotonically strengthens the localization.       

\begin{figure}[htbp]
	\includegraphics[width=1.0\columnwidth,height=0.8\columnwidth]{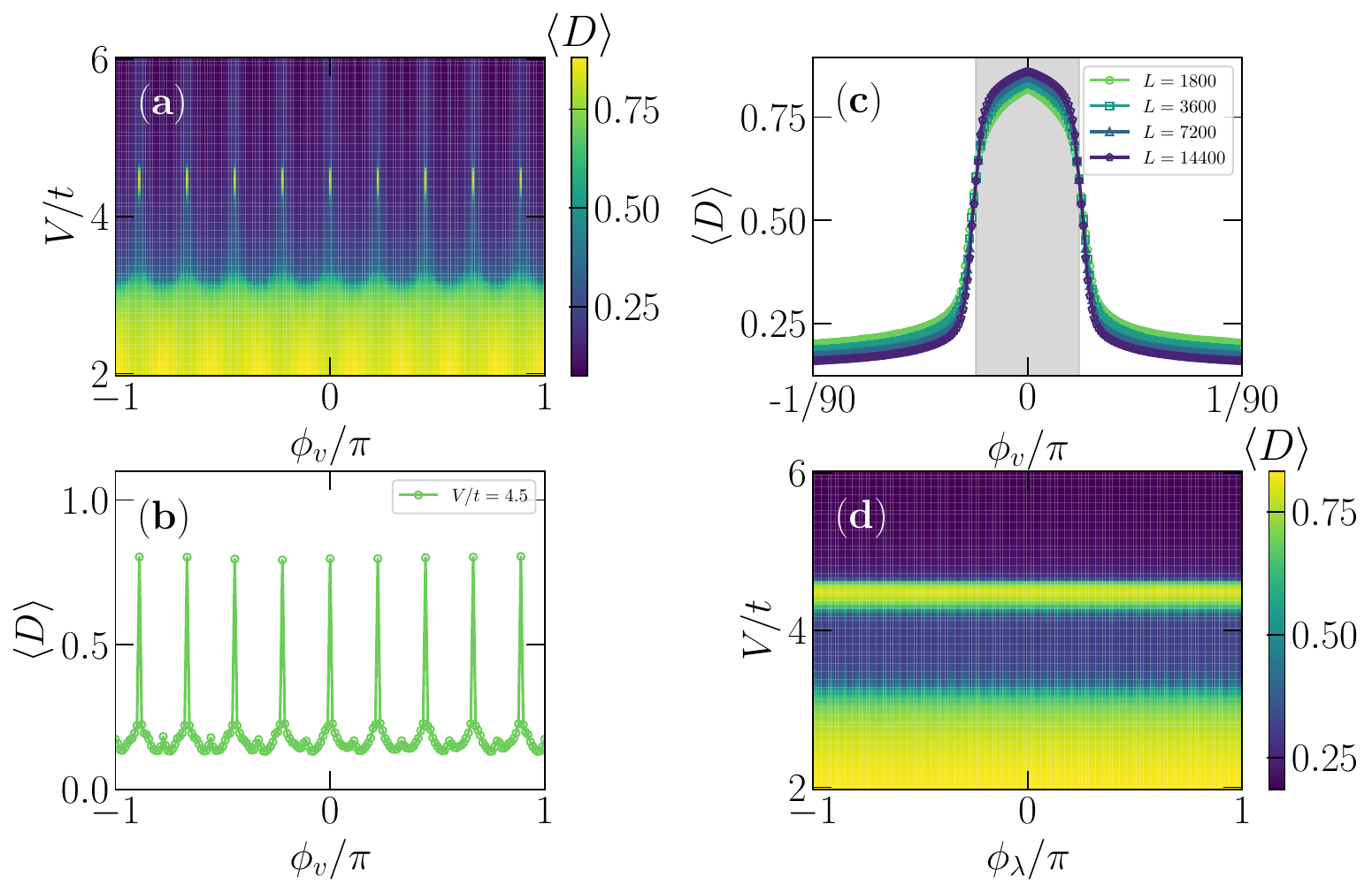}
	\vspace{-0.4cm}
	\caption{(a) Mean fractal dimension $\langle D \rangle$ for different $V/t$ and $\phi_v$ . (b) Mean fractal dimension $\langle D \rangle$ as a function of  $\phi_v$ for $V/t=4.5$. (c) Mean fractal dimension $\langle D \rangle$ for different sizes at $V/t=4.5$. (d) Mean fractal dimension $\langle D \rangle$ for different $V/t$ and $\phi_\lambda$ with $\phi_v=0$.  In (a), (b), and (d), $L=900$ and the step size is $\delta\phi_v/\pi=1/90$. To identify the critical phase factor, a finer step size $\delta\phi_v/\pi=1/9000$ is used in (c). $\phi_\lambda=0$ in (a)-(c). The amplitude of the quasi-periodic potential is set to $\lambda/t=0.007$. 
	}
	\label{fig3}
\end{figure}

Next, we further explore the effect of different phase factors on the reentrant localization transition. Fig.~\ref{fig3} (a) shows the mean fractal dimension $\langle D \rangle$ for various $V/t$ and $\phi_v$. The special case of $\phi_v=0$ has been studied in Fig.~\ref{fig2} (a). By varying $\phi_v$, it is observed that the reentrant localization transition occurs only for specific values of the periodic phase factor $\phi_v$, rather than arbitrarily. 
Moreover, the periodic phase factors associated with scenarios where reentrant localization transitions occur exhibit a clear periodicity, with a period of $\Delta\phi_v/\pi=2/9$. This periodicity is attributed to $\cos (2 \pi \alpha j+\phi_v) =\cos (2 \pi \alpha j+\phi_v+n\Delta\phi_v)$ with $\alpha=1/9$ and integer $n$. In Fig.~\ref{fig3} (b), we show a detailed slice of Fig.~\ref{fig3} (a) for $V/t=4.5$, in which the mean fractal dimension shows peaks at $\phi_v/\pi\approx0, \pm2/9, \pm4/9, \pm6/9,...$, corresponding to extended phases. A more detailed estimate of $\phi_v$ is shown in Appendix~\ref{add5}. To confirm the critical phase factors of the periodic potential, we employ a finer step size to study the mean fractal dimension for different system sizes in Fig.~\ref{fig3} (c). The range of the abscissa is limited to $\phi_v/\pi\in[-1/90,1/90]$ to highlight the transition points. It can be found that the mean fractal dimension increases with the size only in the interval $\phi_v/\pi\in[-1/375, 1/375]$, displaying an extended phase~\cite{Roy214203,Liu012222,Qi224201}. Outside this interval, the fractal dimension decreases with the size, indicating the localized phases~\cite{Roy214203,Liu012222,Qi224201}. It suggests that the reentrant localization transition induced by the composite potential has stringent requirements on the phase factor of the periodic component. In Fig.~\ref{fig3} (d), we fix $\phi_v=0$ and vary $\phi_\lambda$ to study the effect of the quasi-periodic phase factor on the reentrant localization transition. The scan of $\langle D \rangle$ shows that the system consistently exhibits the reentrant localization transition, regardless of $\phi_\lambda$. It demonstrates that the reentrant localization transition is insensitive to the phase factor of the quasi-periodic potential. This aligns with numerous quasi-periodic works, where the phase factor does not change the physical properties of systems~\cite{aubry1980, Ganeshan146601,Li085119, Wang073204,Wang104504, Cai176403,Madeira224505}.

\begin{figure}[htbp]
	\includegraphics[width=1.0\columnwidth,height=0.8\columnwidth]{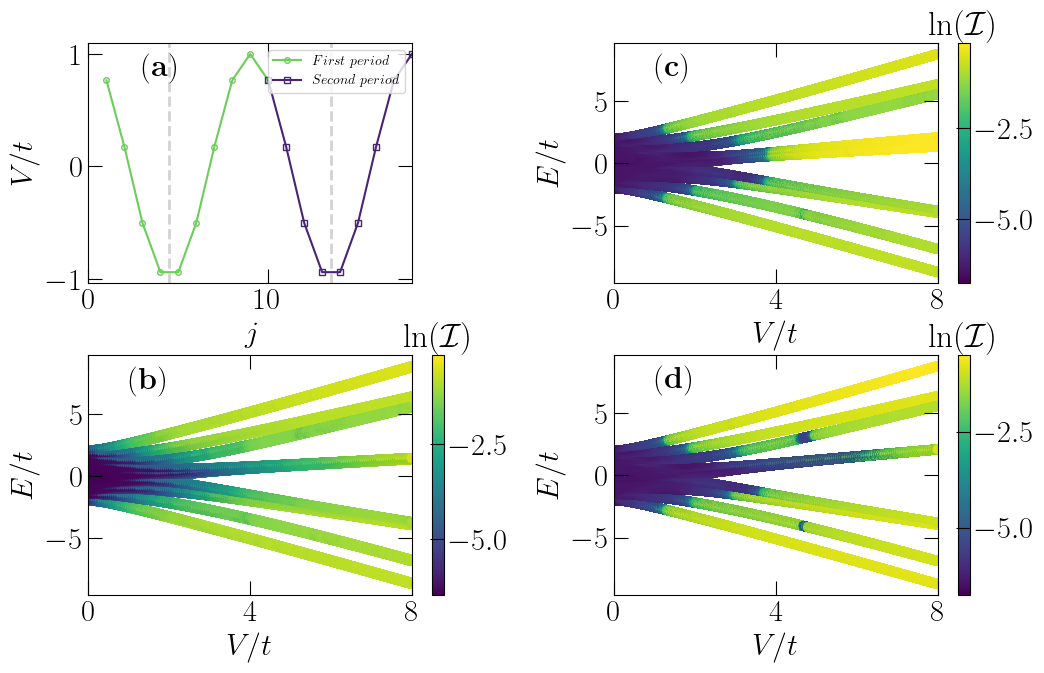}
	\vspace{-0.4cm}
	\caption{(a) The structure of the periodic potential for $\phi_v=0$. (b), (c), and (d) The spectra as a function of $V/t$ under perturbations with different symmetries. Specifically, (b) breaks the translational invariance while preserving the mirror symmetry within each period by adding a perturbation term $\frac{1}{L}\sum_j [\frac{j}{9}]\hat{n}_j$, (c) breaks the mirror symmetry while preserving the translational invariance by adding a perturbation term $\Delta V\sum_{(j\mod9=2)}\hat{n}_j$, and (d) preserves both the mirror symmetry and the translational invariance by adding a perturbation term $\Delta V\sum_{(j\mod9 = 2, 7)}\hat{n}_j$. $L=900$, $\lambda/t=0.007$, $\phi_\lambda=0$, and $\phi_v=0$. The perturbation amplitudes in (c) and (d) are $\Delta V/t = 0.1$. The gray dotted lines in (a) mark the symmetry centers within each period. The colors in (b), (c), and (d) represent the values of $\ln(\mathcal{I})$.
	}
	\label{fig4}
\end{figure}

To analyze how the periodic phase factor influences the reentrant localization transition, we conduct a detailed examination of the periodic potential. In Fig.~\ref{fig4} (a), we show the structure of the periodic potential for $\phi_v=0$, which displays two types of symmetries. The first is the translational invariance and the second is the mirror symmetry. To clarify the roles of these two symmetries, we analyze them individually by adding different perturbation terms to Eq.~\eqref{hub}. In Fig.~\ref{fig4} (b), we use a perturbation term $\frac{1}{L}\sum_j [\frac{j}{9}]\hat{n}_j$ to merely break the translational invariance, where $[\cdot]$ denotes the ceiling function~\cite{Choi022122,Yizhi013322}. It is observed that reentrant localization transitions are disrupted in the spectrum, indicating that the translational invariance facilitates the formation of extended states, thereby promoting the reentrant localization transition. In Fig.~\ref{fig4} (c), we introduce another perturbation term $\Delta V\sum_{(j \mod 9=2)}\hat{n}_j$, where $\Delta V$ is the perturbation amplitude, $j \mod 9$ gives the remainder when $j$ is divided by 9, and the remainder is set to 2 as an example. This perturbation targets the sites satisfying $j \mod 9=2$, which breaks the mirror symmetry within each period while preserving the translational invariance. It is worth noting that only perturbations at sites influencing the second and seventh energy bands can destroy the reentrant localization. Fig.~\ref{fig4} (c) shows that reentrant localization transitions also disappear, highlighting the importance of the mirror symmetry in forming extended states and reentrant localization transitions. Furthermore, it also clarifies that the modification of $\phi_v$ in Fig.~\ref{fig1} and Fig.~\ref{fig3} leads to the destruction of the reentrant localization transition, with the underlying cause being the breaking of the mirror symmetry in the periodic potential. In Fig.~\ref{fig4} (d), we perturb the sites satisfying $j \mod 9=2$ and $j \mod 9=7$ with the same amplitude $\Delta V$, i.e., using a perturbation term $\Delta V\sum_{(j\mod9=2,7)}\hat{n}_j$, which does not break either the translational invariance or the mirror symmetry. It can be found that perturbations ensuring both symmetries do not disrupt the reentrant localization transition.

\begin{figure}[htbp]
	\includegraphics[width=1.0\columnwidth,height=0.8\columnwidth]{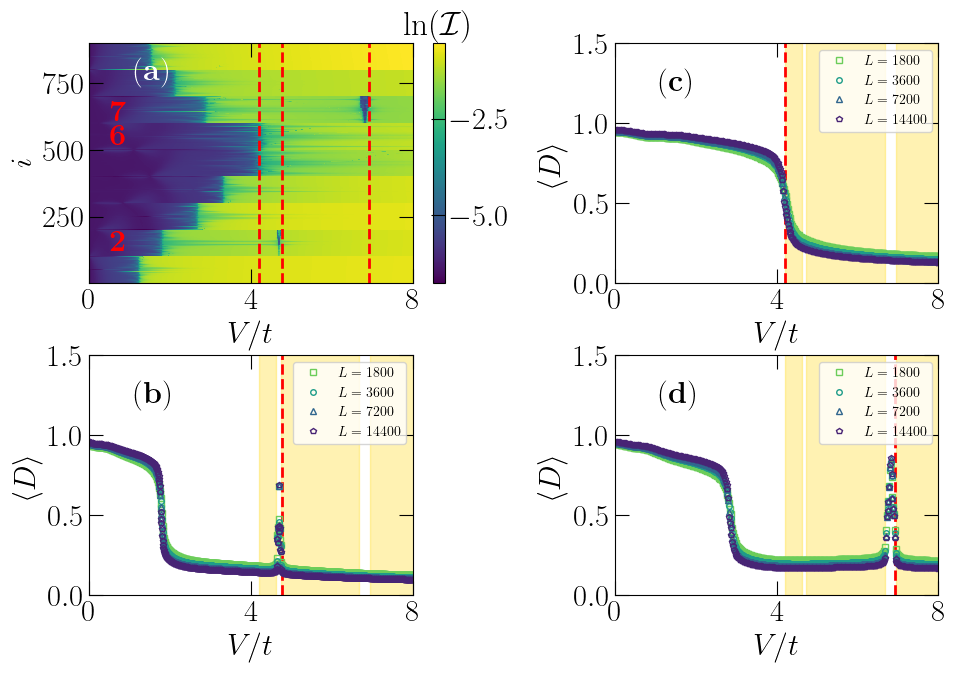}
	\vspace{-0.4cm}
	\caption{(a) The spectrum as a function of $V/t$ for $\alpha=1/9$, perturbed by the perturbation term $\Delta V\sum_{(j\mod9=2,7)}\hat{n}_j$ with $\Delta V/t=0.4$. (b), (c), and (d) The mean fractal dimension $\langle D \rangle$ as a function of $V/t$ for the second, sixth, and seventh bands, respectively. In (a), the ordinate represents the energy level index and colors indicate the values of $\ln(\mathcal{I})$. The red dotted lines mark the localization transition points into the fully localized phases. The yellow-shaded regions in (b), (c), and (d) indicate the fully localized phases, where all states in the spectrum are localized. The system size is $L=14400$ in (a). Other parameters are set to $\lambda/t=0.007$, $\phi_v=0$, and $\phi_\lambda=0$. 
	}
	\label{fig5}
\end{figure}

By adjusting the perturbation amplitude, we find that perturbations preserving spatial symmetries modify the critical amplitude of the reentrant localization transition, which induces fully localized phases, where all states in the spectrum are localized~\cite{RoyChain,Roy214203,Padhanlattice}. Fig.~\ref{fig5} (a) shows the spectrum as a function of $V/t$, which is affected by the perturbation term $\Delta V\sum_{(j\mod9=2,7)}\hat{n}_j$ with a larger amplitude of $\Delta V/t=0.4$ compared to Fig.~\ref{fig4} (d). As $V/t$ increases, the system undergoes three fully localized phases, with their specific locations indicated by the yellow-shaded regions shown in Fig.~\ref{fig5} (b)-(d). To highlight the transition points, we mark them with red dotted lines in Fig.~\ref{fig5} (a), which are a collection of red dotted lines shown in Fig.~\ref{fig5} (b)-(d). It can be found that the transition into the first fully localized phase coincides with the localization transition point of the sixth energy band (The energy band numbers are the same as those in Fig.~\ref{fig1}, and the second, sixth, and seventh energy bands are marked). The first fully localized phase is destroyed when extended states emerge in the second energy band. As these extended states disappear, the second energy band re-enters the localization, leading the system into the second fully localized phase. Analogously, the second fully localized phase is broken when the seventh energy band begins to exhibit extended states. While the reentrant localization transition occurs in the seventh energy band, the system enters the third fully localized phase. It is worth pointing out that, for a certain energy band, such as the second or the seventh energy band, the reentrant localization transition occurs only once as the amplitude increases. The multiple entries into the fully localized phases~\cite{Padhanlattice} are attributed to the misalignment of reentrant localization transitions across different energy bands. To illustrate transitions more intuitively, we present the mean fractal dimension $\langle D \rangle$ as a function of $V/t$ for the second, sixth, and seventh energy bands in Fig.~\ref{fig5} (b)-(d). As $V/t$ increases, the system first enters the fully localized phase at $V/t\approx 4.21$, corresponding to the localization transition point in the sixth energy band in Fig.~\ref{fig5} (c). Further increasing $V/t$, the second transition into the fully localized phase occurs at $V/t\approx 4.77$, which arises from the reentrant localization transition in the second energy band in Fig.~\ref{fig5} (b), and the last localization transition is at $V/t\approx 6.94$ driven by the reentrant localization transition in the seventh energy band in Fig.~\ref{fig5} (d). 

\begin{figure}[htbp]
	\includegraphics[width=1.0\columnwidth,height=0.8\columnwidth]{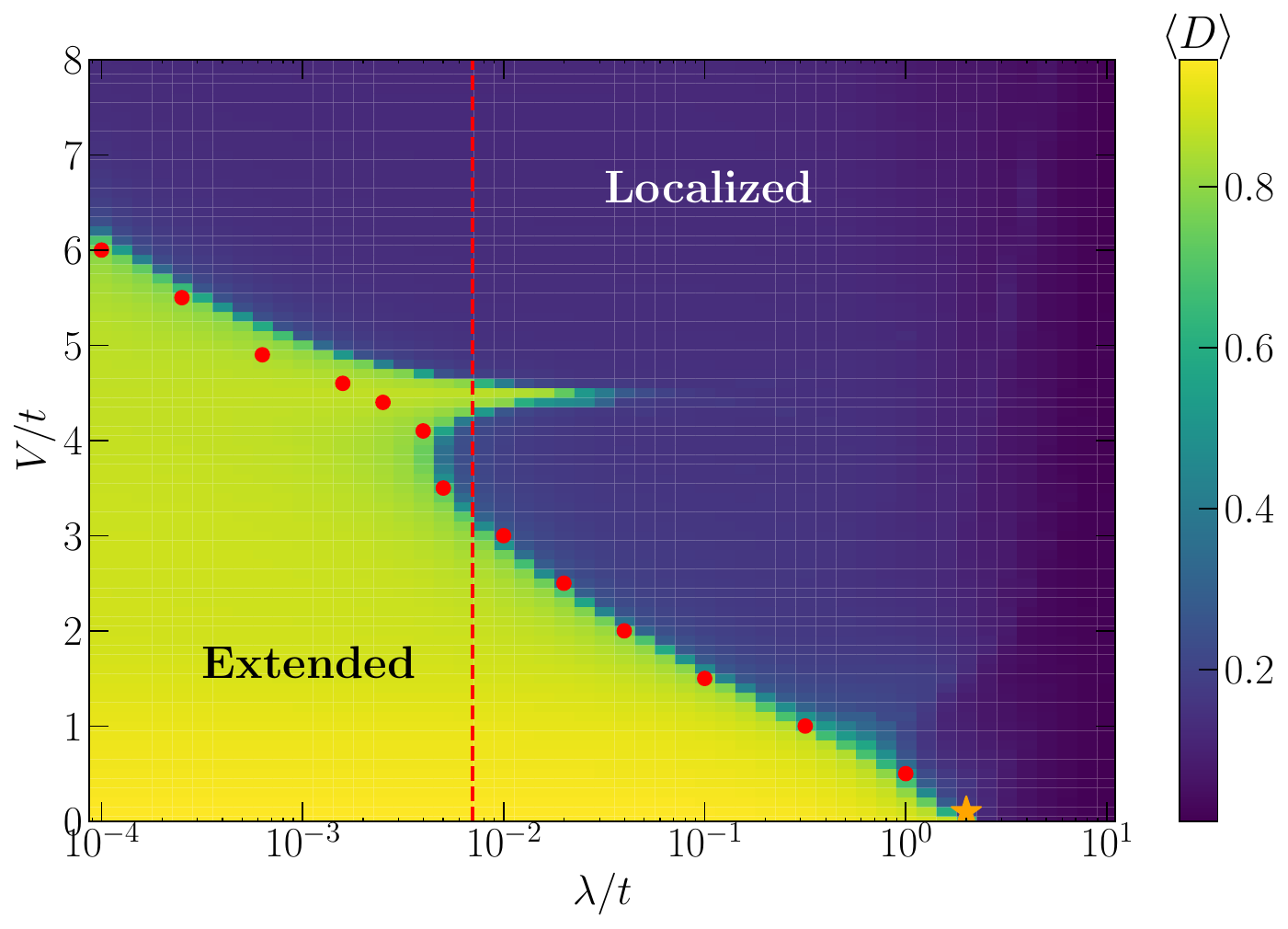}
	\vspace{-0.4cm}
	\caption{Phase diagram of the seventh energy band. The colors indicate the values of $\langle D \rangle$ for $L=14400$ and $\phi_v=\phi_\lambda=0$. The red dotted line marks $\lambda/t=0.007$, which has been studied in Fig.~\ref{fig2}. The orange star indicates the localization transition point of the AA model for $\lambda/t=2$ and $V/t=0$. Red dots mark the transition points between the extended phase and the localized phase for $\phi_v=0.01\pi$.
	}
	\label{fig6}
\end{figure}

Finally, to determine the survival space of the reentrant localization transition, we plot a phase diagram affected by $V/t$ and $\lambda/t$ in Fig.~\ref{fig6}. Here we still take the seventh energy band as an example and fix $\phi_v=0$. The red dotted line highlights the reentrant localization transition for $\lambda/t=0.007$ studied in Fig.~\ref{fig2} (a). In the limiting case of $V/t=0$, the Hamiltonian reduces to the AA model. As the quasi-periodic potential amplitude $\lambda/t$ increases, the system undergoes an extended-localized transition and the transition point is at $\lambda/t=2$~\cite{aubry1980}, which is marked by an orange star in the phase diagram. In another limiting case of $\lambda/t=0$, the composite potential becomes purely periodic, any periodic potential amplitude cannot induce the system to form the Anderson localization~\cite{Abanin021001}. For $V, \lambda \neq 0$, the periodic potential and the quasi-periodic potential compensate each other, that is, as the amplitude of the periodic potential increases, the system requires a smaller amplitude of the quasi-periodic potential to induce the Anderson localization. For comparison, we also plot the transition points between the extended phase and the localized phase for $\phi_v=0.01\pi$ in Fig.~\ref{fig6}, which are marked by red dots. It is obvious that $\phi_v=0$ allows a larger parameter region for the extended phase compared to $\phi_v=0.01\pi$. Notably, whether $\phi_v=0$ or $\phi_v=0.01\pi$, the periodic potential remains translationally invariant. The key distinction between these two scenarios is that the case of $\phi_v=0$ also exhibits a mirror symmetry in the periodic potential. This mirror symmetry contributes to an additional extended phase, leading to the reentrant localization transition. Therefore, a phenomenological explanation for the reentrant localization transition can be described as follows: As $V/t$ increases, the amplitude of the composite potential grows, driving the first localization transition. With further increases in $V/t$, the periodic potential becomes the dominant component of the composite potential.  The mirror symmetry of the periodic potential facilitates the emergence of extended states, resulting in the formation of an extended phase. 
As $V/t$ continues to increase, the periodic component remains dominant. However, the increase in its amplitude leads to a corresponding rise in the amplitude of the composite potential. The quasi-periodic nature and large amplitude of the composite potential promote localized states, resulting in a second localization transition.
In the limiting case for $V/t\rightarrow\infty$, it is expected that the system is in an extended phase. However, in the parameter range adopted in our phase diagram, we do not observe an extended phase for large $V/t$. This is because the quasi-periodic component effectively destroys the translational invariance of the periodic component.

\section{conclusion}
In summary, we investigate the reentrant localization transition induced by a composite potential in a one-dimensional system. Our findings numerically reveal that the reentrant localization transition is not confined to the case of a staggered potential. Instead, this phenomenon occurs in a broader range of systems with composite potentials. Compared to the staggered potential with $\alpha = 1/2$, reentrant localization transitions induced by other periodic potentials, e.g., $\alpha = 1/9$, impose stricter conditions on the implementation parameters. Specifically, the periodic potential should not only exhibit the translational invariance but also possess the mirror symmetry. By introducing different forms of perturbations in the periodic potential, we confirm that violating either the translational invariance or the mirror symmetry alone may lead to the disappearance of the reentrant localization transition, whereas perturbations preserving both symmetries do not destroy it. Furthermore, we uncover that multiple localization transitions into fully localized phases can be regulated by the misalignment of reentrant localization transitions across different energy bands. In the end, by utilizing a global phase diagram, we reveal that the reentrant localization transition induced by the composite potential is driven by the mirror symmetry of the periodic potential.

\begin{acknowledgments}
We acknowledge support from the Natural Science Foundation of China (Grant No. 12404322),
the Zhejiang Provincial Natural Science Foundation of China under Grant No. LQ24A040004,
and the Science Foundation of Zhejiang Sci-Tech University (Grant No. 23062152-Y). 
\end{acknowledgments}

\section*{Data Availability Statement} 
The data that support the findings of this article are openly available ~\cite{data}.

\bibliography{reference}

\clearpage

\onecolumngrid

\appendix
\begin{appendices}
	
\vspace{0.3cm}
	
\twocolumngrid
	
\beginsupplement
	
\section{Reentrant localization transition for different $\alpha$ }~\label{add1}
\begin{figure}[htbp]
	\includegraphics[width=1.0\columnwidth,height=0.8\columnwidth]{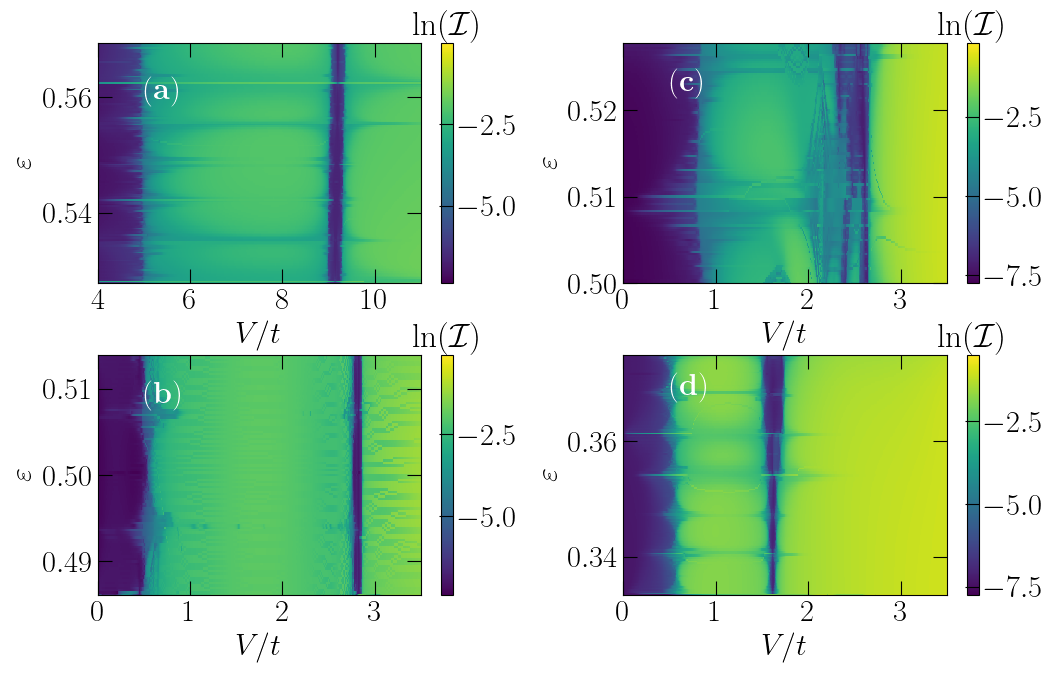}
	\vspace{-0.4cm}
	\caption{(a)-(d) The spectra as a function of $V/t$ for $\alpha=1/3$, $\alpha=1/4$, $\alpha=1/5$, and $\alpha=1/6$, respectively. The colors denote the values of $\ln(\mathcal{I})$. The ordinate represents the normalized energy level index $\varepsilon=i/L$. To illustrate the reentrant localization transition clearly, only a portion of the spectrum is displayed. $L=3600$, $\beta=(\sqrt{5}-1)/2$, $\phi_\lambda=0$. The implementation of the reentrant localization transition is affected by $\phi_v$ and $\lambda$. We set $\phi_v=0$ and $\lambda=10^{-0.4}$ in (a),  $\phi_v=0$ and $\lambda=10^{0.2}$ in (b), $\phi_v=0$ and $\lambda=1$ in (c), $\phi_v=\pi/2$ and $\lambda=10^{-0.1}$ in (d). 
		}
	\label{figadd1}
\end{figure}

\begin{figure}[htbp]
	\includegraphics[width=1.0\columnwidth,height=0.8\columnwidth]{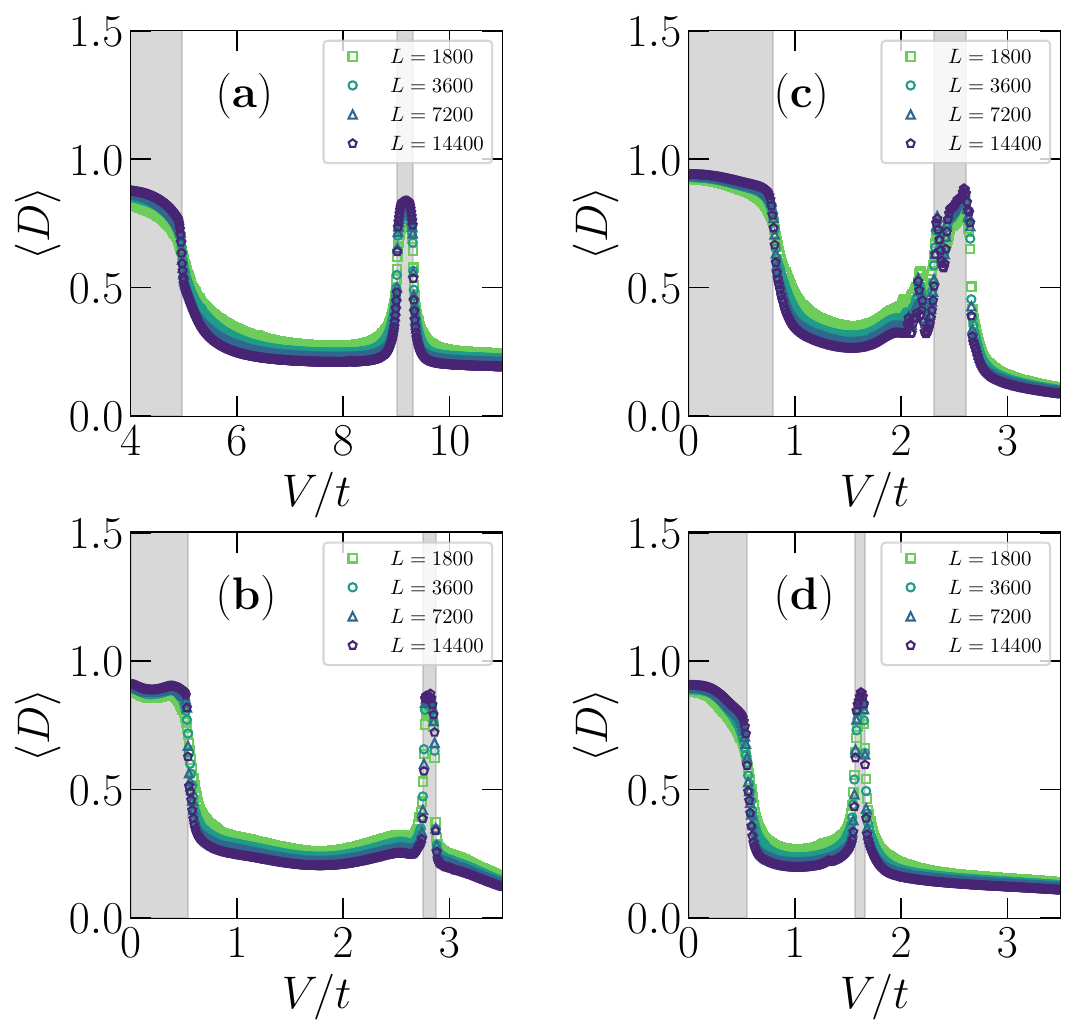}
	\vspace{-0.4cm}
	\caption{(a)-(d) Mean fractal dimension $\langle D \rangle$ as a a function of $V/t$ for $\alpha=1/3$, $\alpha=1/4$, $\alpha=1/5$, and $\alpha=1/6$, respectively. The ordinate $\langle D \rangle$ represents the mean fractal dimension of 20 states near the target normalized energy level index $\varepsilon$. In (a)-(d), $\varepsilon=0.55$, $\varepsilon=0.50$, $\varepsilon=0.52$, and $\varepsilon=0.35$, respectively. Other parameters are the same as those in Fig.~\ref{fig1}. The gray-shaded regions indicate the extended phases.
		}
	\label{figadd2}
\end{figure}

In Fig.~\ref{figadd1}, we show the spectra as a function of $V/t$ for $\alpha=1/3$, $1/4$, $1/5$, and $1/6$. To clearly demonstrate the reentrant localization transition, we focus on a specific part of the spectrum for detailed analysis. The ordinate is the normalized energy level index $\varepsilon=i/L$, where $i$ is the energy level index. The periodic phase factors $\phi_v$ used in Fig.~\ref{figadd1} (a)-(d) ensure that the periodic potential retains both the translation invariance and the mirror symmetry. As the amplitude of the periodic potential $V/t$ increases, the system undergoes a sequence of extended-localized-extended-localized phases in all four cases of $\alpha=1/3$, $1/4$, $1/5$, and $1/6$. It indicates the prevalence of the reentrant localization transition in the systems subjected to a composite potential. For different rational wave vectors, the critical amplitudes and positions in the spectrum where the reentrant localization transition occurs are different. To eliminate the impact of finite-size effects, we calculate the mean fractal dimension $\langle D \rangle$ for different system sizes, as shown in Fig.~\ref{figadd2}. In the extended phases, which are marked by gray-shaded regions, $\langle D \rangle$ increases with the size. Conversely, in the localized phases, $\langle D \rangle$ decreases as the size grows. The transition points indicated in Fig.~\ref{figadd2} are consistent with those shown in Fig.~\ref{figadd1}.

\begin{figure}[htbp]
	\includegraphics[width=1.0\columnwidth,height=0.8\columnwidth]{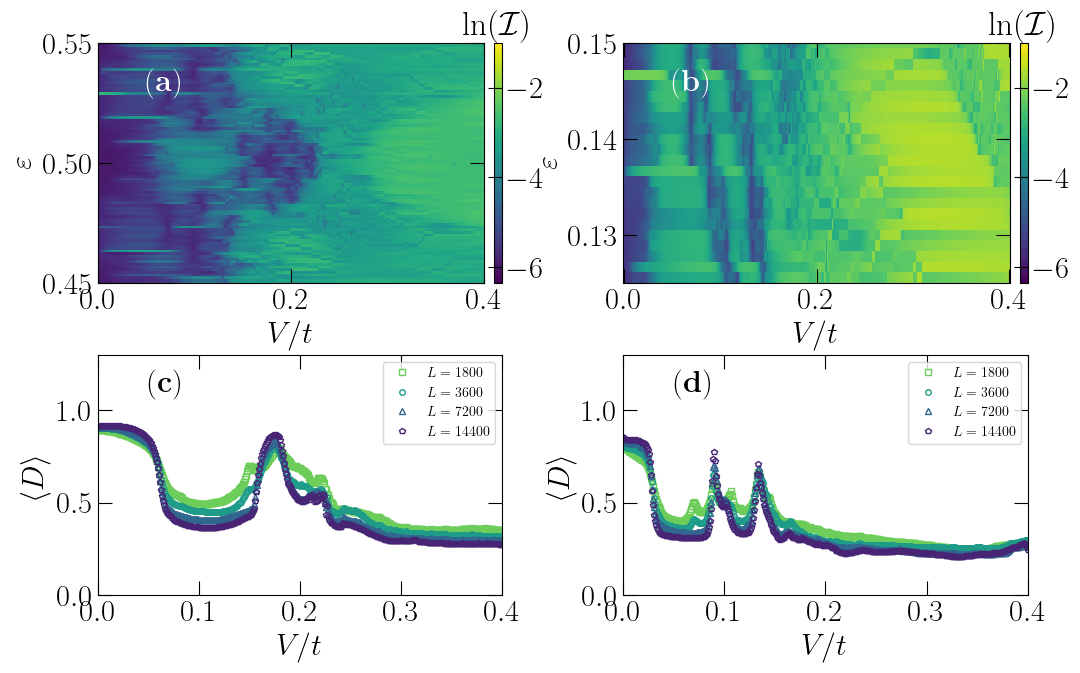}
	\vspace{-0.4cm}
	\caption{(a) and (b) The spectra as a function of $V/t$ for a large denominator in $\alpha$ near $\varepsilon\approx 0.5$ and $\varepsilon\approx 0.14$.  (c) and (d) Mean fractal dimension $\langle D \rangle$ as a a function of $V/t$ for different system sizes, corresponding to (a) and (b). $L=900$ in  (a) and (b). $\lambda/t=1.5$, $\alpha=1/100$, $\beta=(\sqrt{5}-1)/2$, and $\phi_\lambda=\phi_v=0$.
	}
	\label{figadd5}
\end{figure}

In addition to the cases of small denominators in $\alpha$ above, we also investigate systems with large denominators, e.g., $\alpha=1/100$. In Fig.~\ref{figadd5} (a) and (b), we present the spectra as a function of $V/t$ and zoom in on the details at $\varepsilon\approx 0.5$ and $\varepsilon\approx 0.14$.  As $V/t$ increases, the system undergoes two localization transitions for $\varepsilon\approx 0.5$, whereas the system undergoes three localization transitions for $\varepsilon\approx 0.14$. This further supports the conclusion in the main text that the reentrant localization
transition is universal in systems with composite potentials. In Fig.~\ref{figadd5} (c) and (d), we study the mean fractal dimensions $\langle D \rangle$ for different system sizes, which correspond to Fig.~\ref{figadd5} (a) and (b), respectively.
It can be observed that the mean fractal dimension increases with system size in the extended phases, whereas it decreases with system size in the localized phases. By carefully examining the details of the band structure, we find that the emergence of multiple reentrant extended phases results from the various overlaps of subbands as $V/t$ increases.

\section{Lyapunov exponent}~\label{add2}
To calculate the Lyapunov exponent, we first rewrite Eq. \eqref{hub} in the form of a Schr\"{o}dinger equation
\begin{equation}\label{h}
	-t\left(\psi_{j+1}+\psi_{j-1}\right)+V_j\psi_{j}=E\psi_{j},
\end{equation}
where $\psi_{j}$ denotes the amplitude of the wavefunction at site $j$ and $E$ is the eigenvalue. The composite potential is given by 
\begin{equation}
	V_j=V \cos (2 \pi \alpha j+\phi_v)+\lambda \cos (2 \pi \beta j+\phi_{\lambda}).
\end{equation}
We then transform this equation into the transfer matrix form
\begin{equation}\label{matrix}
	\left[\begin{array}{c}
		\psi_{j+1} \\
		\psi_j
	\end{array}\right]=T_j\left[\begin{array}{c}
		\psi_j \\
		\psi_{j-1}
	\end{array}\right],
\end{equation}
where the transfer matrix $T_j$ reads
\begin{equation}\label{T_j}
	\begin{aligned}
		T_j & =\left(\begin{array}{cc}
			E-V_j & -1 \\
			1 & 0
		\end{array}\right)\\
		  &= \left(\begin{array}{cc}
		  	E-V \cos (2 \pi \alpha j+\phi_v) \\
		  	-\lambda \cos (2 \pi \beta j+\phi_{\lambda}) & -1 \\
		  	1 & 0
		  \end{array}\right), 
	\end{aligned}
\end{equation}
here $t=1$ is used as the energy unit. The Lyapunov exponent measures the exponential rate of growth of the transfer matrix product, which is defined as
\begin{equation}\label{e5}
	\begin{aligned}
		\gamma(E) &=\lim _{L \rightarrow \infty} \frac{1}{L} \ln (||\prod_{j=1}^{L} T_j||).
	\end{aligned}
\end{equation}
In this work, we numerically calculate the Lyapunov exponent. For the calculation details, we refer to Ref.~\cite{Wei3314}.

\section{Critical amplitudes and exponents}~\label{add3}
\begin{figure}[htbp]
	\includegraphics[width=1.0\columnwidth,height=0.8\columnwidth]{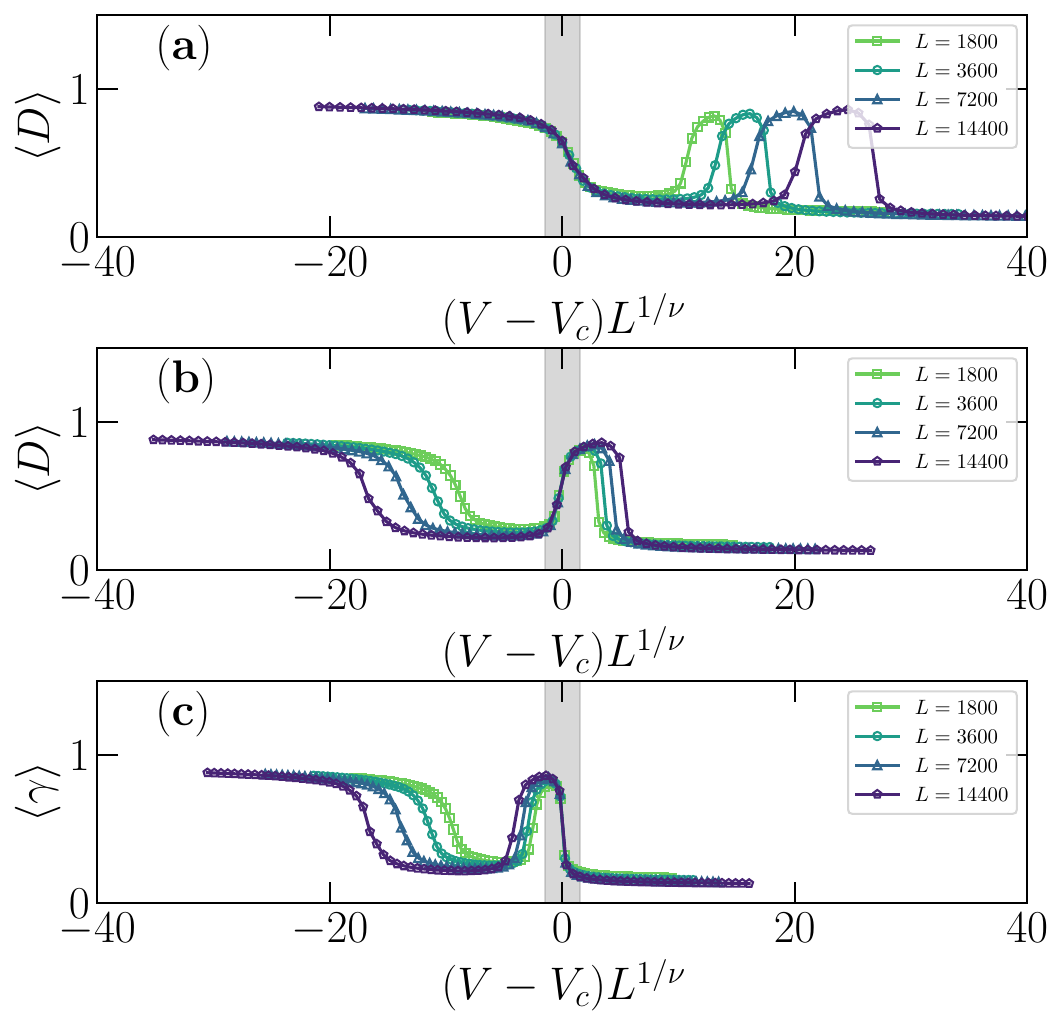}
	\vspace{-0.4cm}
	\caption{(a)-(c) Mean fractal dimension $\langle D \rangle$ as a function of $(V-V_c)L^{1/\nu}$ for $V_c=3.15$, $V_c=4.28$, $V_c=4.62$, respectively. $\nu=3.3$, $3.5$, and $3.9$ in (a)-(c). Parameters are the same as those in Fig.~\ref{fig2} (a) in the main text. The gray-shaded regions indicate the fitting regions near $(V-V_c)L^{1/\nu}\approx 0$.
	}
	\label{figadd3}
\end{figure}

In Fig.~\ref{figadd3}, we use the ansatz $f[(V-V_c)L^{1/\nu}]$~\cite{Luitz081103,Lv013315} to conduct finite-size scaling analysis. Panels (a)-(c) of Fig.~\ref{figadd3} illustrate the first extended-localized transition, the localized-extended transition, and the second extended-localized transition as shown in Fig.~\ref{fig2} (a). By fitting data from different system sizes, we determine the critical amplitudes to be $V_c=3.15$, $V_c=4.28$, and $V_c=4.62$, respectively. The corresponding critical exponents are $\nu=3.3$, $3.5$, and $3.9$.

\section{Density of states}~\label{add4}
\begin{figure}[htbp]
	\includegraphics[width=1.0\columnwidth,height=0.8\columnwidth]{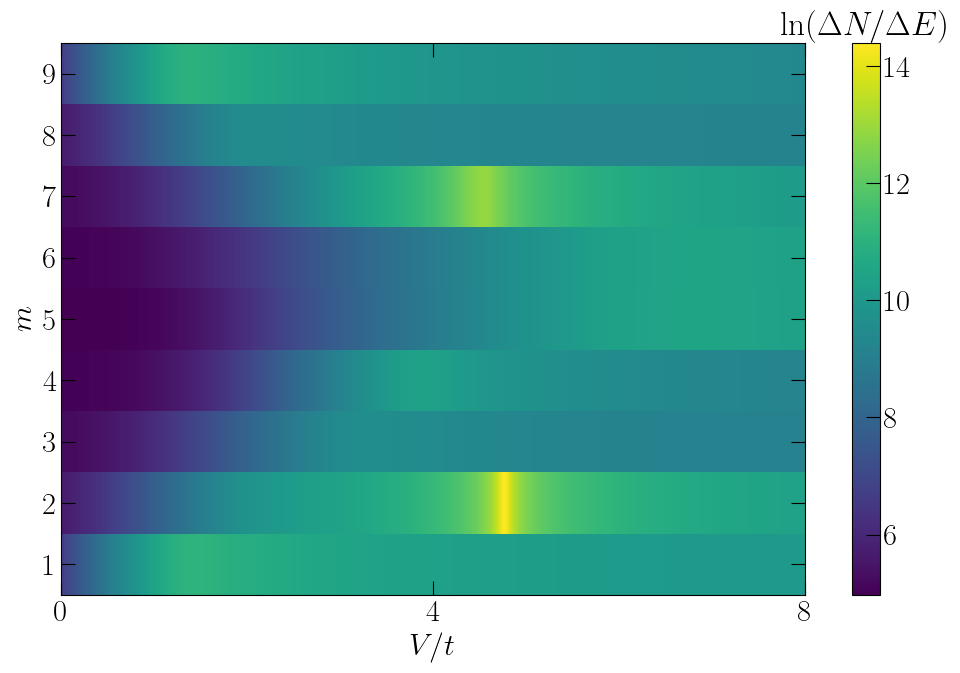}
	\vspace{-0.4cm}
	\caption{Density of states for different subbands. The ordinate $m$ indicates the energy band index. The parameters are the same as those in Fig.1 (b) in the main text. The colors represent the values of $\ln(\Delta N/\Delta E)$, where $\Delta N$ is the number of states within the subband, and $\Delta E$ is the bandwidth of the subband.
	}
	\label{figadd4}
\end{figure}

In Fig.~\ref{fig1} (a) and (b), we show that only the second and seventh energy bands exhibit the reentrant localization transition. This is because only these two subbands have narrow bandwidths with a high density of states under certain parameters, which is conducive to the overlap of localized states to form extended states, thereby enabling the reentrant localization transition. An intuitive demonstration is shown in Figure~\ref{figadd4}. As $V/t$ increases, the second (seventh) energy band shows a relatively high density of states for $4.70<V/t<4.81$ ($4.28<V/t<4.62$), corresponding to the region where the system re-enters the extended phase after the first localization transition in Fig.~\ref{fig1} in the main text. Analogically, for other values of $\alpha$ in Fig.~\ref{figadd1}, it can also be observed that a relatively high density of states is present in the reentrant extended phase after the first localization transition.

\section{Estimate singular values of $\phi_v$}~\label{add5}
\begin{figure}[htbp]
	\centering
	\includegraphics[width=1\columnwidth,height=0.8\columnwidth]{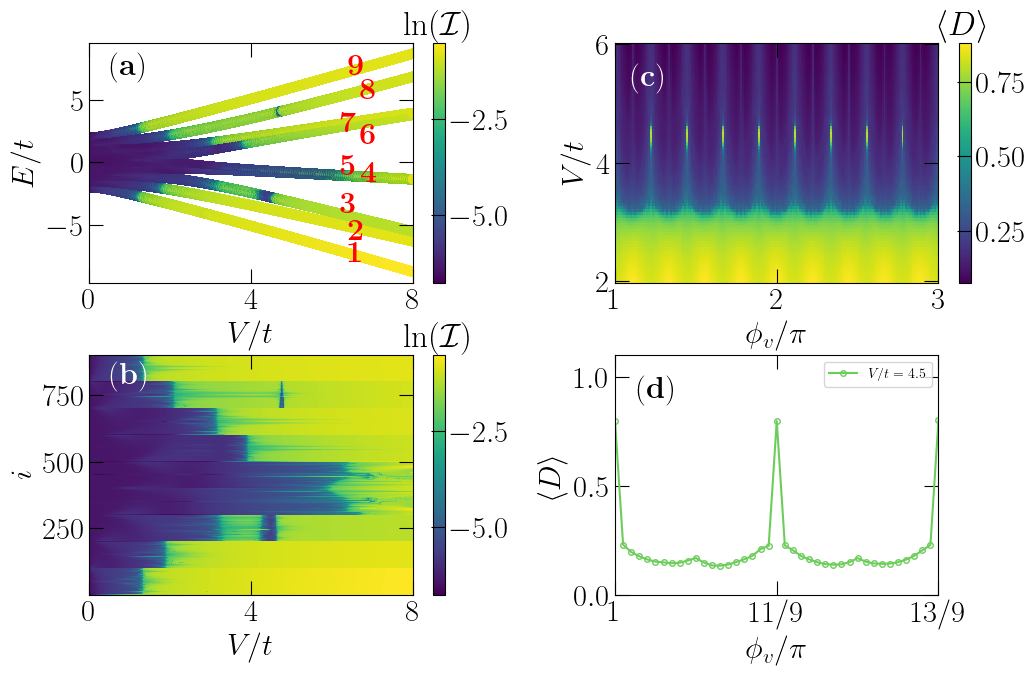}
	\vspace{-0.6cm}
	\caption{(a) and (b) The spectra as a function of $V/t$ for $\phi_v=\pi$. (c) Mean fractal dimension $\langle D \rangle$ for different $V/t$ and $\phi_v$ in the third energy band. (d) The detailed slice of (c) for $V/t=4.5$. $L=900$, $\lambda/t=0.007$, and $\phi_\lambda=0$. }
	\label{figadd6}
\end{figure}

In the main text, we demonstrate that $\phi_v$ influences reentrant localization by altering the symmetry of the periodic potential. Therefore, we can estimate the singular values of $\phi_v$ through a symmetry analysis of the periodic potential (the range of $\phi_v$ is very narrow, which allows us to regard it as an effectively singular). Here, we continue to set $\alpha=1/9$, and first consider the mirror symmetry
\begin{equation}
	\begin{aligned}
		cos(2\pi\frac{1}{9}j+\phi)=cos[2\pi\frac{1}{9}(9-j)+\phi].
	\end{aligned}
\end{equation}
Utilizing the property of the cosine function, $cos(A)=cos(B)$ implies $A=\pm B+2\pi k$, where $k$ is an integer, we obtain
\begin{equation}
	\begin{aligned}
		2\pi\frac{1}{9}j+\phi = \pm[2\pi\frac{1}{9}(9-j)+\phi]+2\pi k.
	\end{aligned}
\end{equation}
Simplifying this equation yields
\begin{equation}
	\begin{aligned}
		\phi = \pi (k -1).
	\end{aligned}
\end{equation}
Next, we consider the translational invariance. Assume that translating the system by $m$ sites results in a phase shift of 
$\phi'$. This condition implies
\begin{equation}
	\begin{aligned}
		cos(2\pi\frac{1}{9}(j+m)+\phi)=cos(2\pi\frac{1}{9}j+\phi+\phi').
	\end{aligned}
\end{equation}
From this relation, we obtain
\begin{equation}
	\begin{aligned}
		\phi'=\frac{2\pi}{9} m, \qquad m=0, \pm 1, \pm 2, .......
	\end{aligned}
\end{equation}
By combining the results for the mirror symmetry condition ($\phi$) and the translational invariance condition ($\phi'$), the possible values of $\phi_v$ are given by
\begin{equation}
	\phi_v = \begin{cases}
		\frac{2n}{9} \pi, \\
		(\frac{2n}{9}+1) \pi,
	\end{cases}
\end{equation}
where $n=0, \pm 1, \pm 2, .......$ is an integer. In the main text, we have shown that the reentrant localization transition exists in the case of $\phi_v=\frac{2n}{9} \pi$. Here, we also provide the cases of $\phi_v=(\frac{2n}{9}+1)\pi$. First, we present the spectra as a function of $V/t$ for $\phi_v = \pi$ in Figure~\ref{figadd6} (a) and (b), corresponding to $n=0$. From low energy to high energy, reentrant localization transitions occur in the third and eighth energy bands. In Figure~\ref{figadd6} (c), we focus on the third energy band and present the mean fractal dimension $\langle D \rangle$ as a function of $V/t$ and $\phi_v$. The results reveal that the reentrant localization transition occurs periodically for $\phi_v = \left(\frac{2n}{9} + 1 \right) \pi$. For further clarity, Figure~\ref{figadd6} (d) provides a view of the slice of Figure~\ref{figadd6} (c) for $V/t = 4.5$, which clearly shows that the period interval is $\Delta\phi_v=\frac{2}{9}\pi$.

\end{appendices}
\end{document}